\documentstyle[12pt,aasms4]{article}

\begin{document}
\def\etal{{\it et al.}}
\small

\title{~~~~~~~~Disentangling the Dynamical Mechanisms
\newline for Cluster Galaxy Evolution}

\centerline{Xiaolei Zhang$^{1,2}$}

\centerline{$^{1}$Remote Sensing Division}
\centerline{US Naval Research Laboratory}
\centerline{Washington, DC 20375, USA}

\smallskip

\centerline{$^{2}$ Current Affiliation:}
\centerline{Department of Physics and Astronomy}
\centerline{George Mason University}
\centerline{Fairfax, VA 22030, USA}
\centerline{xzhang5@gmu.edu}

\begin{abstract}

The determination of the dynamical causes of the morphological Butcher-Oemler 
(BO) effect, or the rapid transformation of a large population of late-type 
galaxies to earlier Hubble types in the rich cluster environment between 
intermediate redshifts and the local universe, has been an important 
unsolved problem which is central to our understanding of the general problems 
of galaxy formation and evolution. In this article, we survey the existing
proposed mechanisms for cluster galaxy transformation, and discuss
their relevance and limitations to the explanation of the morphological
BO effect.  A new infrared diagnostic approach is devised to
disentangle the relative importance of several major physical
mechanisms to account for the BO effect, and an example of the first 
application of this procedure to a single rich, intermediate redshift galaxy
cluster is given to demonstrate the viability of this approach.
The preliminary result of this analysis favors the interaction-enhanced
secular evolution process as the major cause of the cluster-galaxy 
morphological transformation. This conclusion is also supported by 
a wide range of other published results which are assembled here
to highlight their implications on a coherent physical origin for the 
morphological BO effect.
\end{abstract}

%\keyword{Galaxies}

\section{Introduction}

Central to the study of the cosmic evolution of galaxies is the identification 
of the dominant dynamical mechanism responsible for the rapid transformation, 
between intermediate redshifts and the local universe, of cluster galaxies 
from a population characterized by a significant fraction of late-type spirals 
to one dominated by S0s and ellipticals: the so-called morphological 
Butcher-Oemler (BO) effect (Butcher \& Oemler 1978a,b; Dressler et al. 1997).

At the heart of the BO effect is the spiral to early-type transformation
(Dressler et al. 1997; Ellis 1998).  This transformation not only involves the 
extinguishment of the star-formation activity, but also the gradual increase 
of bulge-to-disk ratio (Franx 2004; Treu 2004 and the references therein).  
Furthermore, whichever physical mechanism(s) produced the BO effect, in addition 
to producing the observed number density evolution of cluster galaxies, must 
simultaneously be able to preserve the relatively tight color-magnitude relation 
observed throughout the redshift range z=0-1 (van Dokkum \& Franx 2001).  The relevant 
process for galaxy transformation must also preserve the outer disks
and spiral structures observed 
for many of these galaxies during the quenching of star formation and the 
morphological transformation (Couch et al. 1998; Poggianti et al. 1999; 
Goto et al. 2003b).
 
Several candidate mechanisms have been proposed to account for the 
morphological BO effect.  These include: major and 
minor mergers (Icke 1985; Kauffmann 1995), galaxy harassment through tidal 
shocks (Moore et al. 1996,1998,1999), galaxy infall onto the forming cluster and 
the resulting stripping of either the disk or halo gas (with the latter having
the name of ``starvation'') due to the ram pressure of the intracluster medium
(Gunn \& Gott 1972; Larson, Tinsley, \& Caldwell 1980), as well as a
newly-discovered secular dynamical process mediated by the large-amplitude
density wave patterns excited during the tidal interaction of galaxies in
clusters (Zhang 1996,1998,1999; Zhang \& Buta 2007a,b).

The dominant mechanism responsible for producing the BO effect is still 
uncertain.  Most of the previously-proposed mechanisms run into contradictions
with one or another of the observed characteristics of cluster galaxies upon
careful analyses.  Due to the high-speed nature of galaxy-galaxy encounters 
within clusters which prevents the encountering galaxies from ``sticking''
to one another, major mergers are not believed to have a significant 
probability in clusters (Dressler et al. 1997), though an exception to this 
conclusion has been observed for one high-z cluster (van Dokkum et al. 1998).
Moreover, there does not seem to be a large reservoir of dwarf spheroidal 
satellites in these clusters (Trentham 1997) to cause the simultaneous 
morphological transformation of the large number of BO galaxies by minor 
mergers. The tightness of the color-magnitude relation of cluster early-type 
galaxies (Franx 2004 and the references therein) and the well-preserved disks 
throughout the BO evolution process (Moran et al. 2007b and the references
therein) also argue against a significant role played by mergers.

Ram pressure stripping of the interstellar medium for infalling cluster 
galaxies (Gunn \& Gott 1972) by itself appears not to be sufficient to
account for the observed level of cluster galaxy transformation. Stripping of
gas generally does not lead to a significant change in the bulge-to-disk (B/D)
ratio, though Larson et al. (1980) had envisioned the reduction of disk 
size due to gas-stripping as one way of increasing the B/D.
However, recent observations and analysis by Moran et al. (2007a) showed that
cluster spiral galaxies have systematically higher central density
than their field counterparts, an effect which is difficult to produce if
the outer-disk stripping is the major cause of the cluster BO effect.

There is also the issue of timescale with any stripping/starvation mechanism
which relies on the infall of galaxies from the field region into the centers
of clusters: If an infalling galaxy starts its journey from the 
field at 5 Mpc distance to the cluster center, and has an average velocity of 
500 km/sec\footnote{Even though the cluster velocity dispersion 
is generally on the order of 1000 km/sec, the velocities of galaxies at 
the outskirts of the clusters, especially their radial components, are 
generally quite low, since these outer galaxies have not been virialized and thus have not gained
the kind of high velocity dispersion that the cluster-center galaxies have.}, it takes
the galaxy $4 \times 10^9$ years to reach the central region.  This is
a significant fraction of the Hubble time, and comparable to the cluster formation/evolution
timescale.  So the late-type galaxies observed in the central regions of
clusters as found by Dressler et al. (1997) are likely to be primordial 
instead of new in-fallers from the field region.  The existence of the morphology-density relation
over several orders-of-magnitude variation of local surface density (Dressler 1980;
Dressler et al. 1997; Goto et al. 2003c), from the densest region of cluster center all the way to
the group environment in the field, also supports the segregation of the varying-density
regions of a cluster during cluster-galaxy morphological transformation, 
and argues against significant mixing of field populations into the central regions of clusters.
Fujita (2004) proposed the idea that galaxies could be ``pre-processed'' 
in groups and intermediate density regions, which then become mixed with the 
cluster in somewhat less time than the field-to-cluster-core infall
time.  However, adding a pre-processed piece of material onto the outskirts of the cluster
does not have the same effect as mixing thoroughly the material
of this newly-added piece into the entire cluster region including the central
core.  This could have allowed the preservation of the relative identities of the
different cluster regions as the cluster grows and matures, 
as well as the maintenance of the observed morphology-density relation.

The galaxy harassment mechanism (Moore et al. 1996,1998,1999) has gained some
popularity in recent years. When first proposed, harassment was meant as a 
process whereby the gravitational tidal shock from a large cluster galaxy tears away 
the outer disk of a small late-type neighboring galaxy, thus transforming the 
small galaxy into an early-type dwarf spheroidal galaxy. Numerical simulations 
by Moore et al. (1998) showed that the harassment process generates remnants 
which are predominantly prolate in shape, as opposed to the oblate shape that 
an S0 galaxy (which is the typical end-product of BO transformation)
usually has.  The follow-on exploration of the harassment 
mechanism by Gnedin (1999, 2003) showed that this mechanism is rather 
inadequate in transforming the morphology of large disk galaxies: even though
a moderate puffing-up of the entire disk was observed, there is not a
significant increase in the bulge-to-disk ratio which is a required
element for the spiral-to-S0 transformation.  The simulations conducted
by Moore et al. (1999) showed that the low-surface-brightness galaxies
in their tests were transformed to dwarf spheroical galaxies due to
tidal shocks, whereas for high-surface-brightness galaxies ``the scaleheight
of the discs increases substantially and no spiral features remain''.  This
is in marked contrast to the post-interacting galaxies observed by, e.g.
Poggianti et al. (1999) and Moran et al. (2007b), which showed prominent
spiral structures and well-preserved outer disks after whatever interaction
processes which had resulted in the star-burst episodes in these galaxies.

Another proposed mechanism for producing the BO effect is the secular 
dynamical evolution process in galaxies which is further enhanced by 
the tidal interactions in the cluster environment (Zhang 1996,1998,1999; 
Zhang \& Buta 2007a,b).  In this scenario, tidal 
forces between neighboring galaxies and between a galaxy and the cluster 
potential serve to excite large-amplitude, open spiral or bar patterns. Most 
of the morphological transformation is realized through the 
post-interaction mass redistribution process due to the presense of
tidally-enhanced density wave patterns.
In this process the radial mass accretion throughout the galaxy disk 
and the vertical heating of stars occur simultaneously over the time span of a few Gyrs, 
due to the irreversible energy and angular momentum
exchange between the density wave and the disk matter and the outward transport
of these exchanged energy and angular momentum by the density wave to the outer disk,
which lead to the gradual build-up of the bulge 
(Zhang 1999; Zhang \& Buta 2007a,b), in addition to the consumption of gas 
through density-wave induced star formation.  This mechanism has been shown to 
be able to preserve the tight scaling relation of early-type galaxies 
during the secular mass accretion process (Zhang 2004).

\section{An Infrared Diagnostic Approach for the Star-Formation States
of Cluster Galaxies} 

In order to assess the relevance of the different proposed mechanisms
to the explanation of the BO effect, we have developed an infrared (IR)
diagnostic approach which has the potential to shed new light on
the BO problem using the existing and soon-to-be-available data
offered by the new generation of the space IR telescopes
such as Spitzer and Herschel.

Over the past three decades since the initial discovery of the BO effect,
the observational studies of cluster galaxy evolution have been mostly
through ground-based and space-based optical telescopes, owing to their
much superior sensitivity, angular resolution, and area-covering efficiency.
However, it is well known that optical observations suffer from dust
extinction which can shield some of the most intense star-formation
activities in clusters (Geach et al. 2006).  On the other hand, the existing 
IR observations, even those done with the state-of-the-art space IR telescopes, 
are limited in their angular resolution compared to optical observations.
For Spitzer Space Telescope observations of intermediate redshift galaxies, 
for example, essentially only the total flux across the different IR
bands are known for each cluster galaxy, with no further spatial
resolution within a single galaxy.  Therefore, given this state of
the current IR observations, the new method is devised to make use
of only the bolometric lumonisty information (which is contributed
most significantly by the IR bands for starburst galaxies) of cluster
galaxies.

The method makes use of the fact that most of the proposed mechanisms
for BO transformation will involve varying levels of dusty starbusrt,
and furthermore these different mechanisms
have different predicted characteristic IR luminosity levels, as
well as different timescales fot sustaining these
different IR luminosity levels.  Given the known BO transformation
rate (which can be determined from a statistical study of the
galaxy morphologies in rich clusters at the intermediate redshifts, compared to 
a similar study for the local universe), we can infer the roles
of different proposed mechanisms by observing the different frequecies
of occurence of the varying IR-luminous galaxies in the different
regions of intermediate clusters.  This quantitative approach has
major advantages in constraining the importance of the candidate
dynamical mechanisms in transforming cluster galaxy morphologies
compared to the simple-minded counting of the IR sources of
a given luminosity range in an observing epoch, because the timescales
of the operation of these dynamical mechanisms are vastly different,
and a simple counting approach can result in serious misconceptions
of the relative contributions of the different processes, as our
example in the next section will show.

For a cluster region of total galaxy number $N$, if there are $n$ galaxies 
observed to be in a luminosity range characteristic of a type of starburst 
event, we denote the observed frequency for that type of event $f_{event}$ as
\begin{equation}
f_{event}  =  {n \over N}
.
\end{equation}

Assume the duration of the starburst event is $\tau_{event}$ (usually 
believed to be on the order of $10^8$ yr for strong-interaction-type 
events such as merger and harassment, and much longer for gas-stripping 
type events and for post-interaction secular evolution), after 
a cluster evolutionary timescale of $T$ (on the order of a few Gyr, the 
exact number depending on the redshift $z$ of a cluster), we expect a fraction
\begin{equation}
F = f_{event} \cdot {T  \over {\tau_{event}} }
\label{eq:eqfraction}
\end{equation}
of the cluster galaxies to have gone through morphological change due to the
particular type of star-forming event alone.  

The total fraction $F$ for BO transformation between an intermediate
redshift and the local universe, due to the yet-unsettled physical mechanism(s),
is now fairly well determined for rich BO clusters
(see, e.g. van Dokkum \& Frank 2001; Franx 2004), and its values is around
20\%, for example, between $z=0.4$ and $z=0$ (see for example Figure 4a of 
van Dokkum \& Franx 2001,
where at $z=0.4$ the early type fraction is about 0.6, and at $z=0$ the early
type fraction is about 0.8, which means a fraction of 20\% of late-to-early
transformation between $z=0.4$ and $z=0$ had happened).  Since this 20\% conversion
fraction is for cluster as a whole, and in the later calculations we divide
the cluster into three broad regions (post-virialization, active virialization,
and outskirts), we estimate the morphological conversion fraction between z=0.4 and z=0
for the active virialization region to be 30\% and for the cluster outskirts 
about $F=10\%$.  No estimates for the post-virialization region is given
since the IR diagnostics are not available there,
but it is expected that the morphological transformation process is going on at some
level even in that region where the star formation has ceased.

We stress that not all of the mechanisms that can transform galaxies will include
a dusty-starburst phase (for example, one proposed pathway
for morphological transformation not involving a dusty starburst is dry mergers). 
Even for those mechanisms which can potentially invoke a dusty
starburst, this phase will only be prominent when there is sufficient
gas reservior in the galaxy at the observing epoch.  This approach 
thus predicts the maximum possible abundance of dusty starbursts that 
can be attributed to each mechanism, to be compared with that of the observed abundance.
So for the proposed approach to work effectively, we make the implicit assumption that
the BO precursor galaxies are gas-rich galaxies, which seemed
to be a valid one since that was how Butcher and Oemler first discoved
them (i.e., by observing their blue colors due to enhanced star formation 
in a gas-rich environment).  For the later stage
of the evolution, after the gas is consumed, we can still have
the interaction and the associated morphological transformation events
but no star-formation signature.  However, the arguments for the validity
of the fractions estimated by this approach require only that
the relevant galaxies are gas rich at the observing epoch z,
and that the interaction strength of the relevant mechanisms
stay roughly constnat throughout z.  They do not require the
gas fraction to remain the same throughout z.
This is because we are relating the star-forming events at z with the overall morphological
transformation rate F throughout z, and the latter does not require
a constant gas supply.  So in this sense the star-formation events
at the observation epoch z is only a diagnostic signature of the underlying
morphological transformation process.

We therefore cannot rule out the possibility that there are other
processes which potentially could be transforming galaxies in a way
that does not generate an IR-bright phase, rather we are ruling out these
mechanisms as the explanation for the observed incidence of dusty
starbursts in those clusters where abundant starbursts are observed. 
In \S 3, we will show that the starbursts which are observed in
one particular intermediate redshift cluster can
essentially {\em only} be explained via the dynamical
mechanism of interaction-enhanced secular evolution, which
demonstrated the power of this approach.

On the other hand, using the same principle this kind of ``frequency test'' can in fact be used
also for sources observed in other wave bands (such as optical), as long as
there is a clearly-identifiable spectral/photometric feature which
has a known timescale of operation.  

If the timescale for the individual
interaction event  $\tau_{event}$ is known, we can invert the
above expression to calculate the appropriate value of $f_{event}$
for each type of interaction event in the appropriate luminosity range and
for the relevant cluster region.  These calculated results are
tabulated in Table 1 for a hypothetical BO cluster redshift 
of $z \sim$ 0.4.  In this table  we list calculated values of $f_{event}$ for each of 
the proposed physical mechanisms we consider, split into several radial
cluster regions, and for transforming galaxies that fall into several
different classes of IR luminosity. We will describe in more detail in the following subsections
of how the frequencies in Table 1 are obtained.

Note that if the predictions/observations
were made at a different $z$ (say $z=0.2$ instead of $z=0.4$), then the
observed fraction of IR sources are expected to be lower due to the
smaller amount of available gas (i.e. some of the galaxies have
already completed the late-type to S0 converson between $z=0.2$ and
z=0.4). This effect is partly taken into account by the outward migration
of the boundaries of the active virializing region with decreasing z,
partly by the corresponding change of $F(z)$ at the new observing epoch z.
Of course part of the $F(z)$ dependence accounts for the reduced elapsed
time for a smaller z as well.  Yet the argument should hold no matter
what observation epoch one uses, as long as one concentrates to the
regions where virializating is going on.  The gradually-enlarging
central part of the cluster then falls into the N/A no-prediction zone.

Note also that in Table 1 we did not break the IR luminosities of the 
moderately-luminous galaxies at the usual LIRG (Luminous InfraRed Galaxies) 
level (which has the lowest luminosity value of $L_{IR} = 10^{11} L_{\odot}$), but rather
at $4 \times 10^{11} L_{\odot}$.  This choice of division is due to our 
preliminary examination of the results of Geach et al. (2006)'s
Spitzer 24 $\mu m$ observations of cluster CL 0024+16 at $z=0.39$.  In this
cluster the majority of the IR sources detected have colors 
intermediate between star-forming and passive galaxies, indicating 
that this is possibly a transitional-type population, and these
transition-type galaxies have luminosities extending to about 
$4 \times 10^{11} L_{\odot}$.

Furthermore, in Table 1 we have divided each cluster into three
approximate zones (Goto et al. 2003a): 
(1) The very central post-virialization region, where the star-formation
activity has ceased by the epoch of observation.  For this region
we do not give predictions of IR bright sources, since we assume these
are rare due to the exhaustion of the gas for galaxies in this region.
However, some post-interaction morphological transformation 
could still be going on here after the extinguishment of star formation.
(2) The  active-virialization region.  For this region the 
virialization-triggered star-formation activity is in full swing,
as will be the star-forming activities due a number of proposed mechanisms.
For clusters at different redshift ranges, this intermediate zone has
been found to propagate from the inner cluster toward outer cluster
region as redshift decreases.  (3) The cluster outskirts.  In this
region the star formation activity is elevated compared to the field,
but not as vigorous as in the active virialization region.

The predictions for the frequencies of IR bright sources are thus given
only for the latter two regions of clusters, since for the first region
all of the proposed mechanisms can be made compatible potentially with
the observation of a lack of star-formation activity there.  The
proposed observed fractions for regions (2) and (3) are further divided
into sub-categories for each proposed dynamical mechanism (assuming it 
operates alone, or is the dominating one in a particular region).

\subsection{Galaxy Merger}

In the popular cold dark matter (CDM) paradigm, galaxy mergers are the
preferred means of morphological evolution of galaxies in clusters (see, e.g.,
Kauffmann 1995). Even though the likelihood of mergers in the dense
regions of clusters has been questioned because of the encounter-speed
arguments, the issue of to what extent mergers play a role in
cluster galaxy evolution, in both the dense regions as well as in
the outskirts, have never been firmly settled, especially due to
the paucity of high angular resolution mid- and far-IR observations
of distant clusters before the launch of Spitzer Space Telescope.

For major mergers between large, gas-rich disk galaxies, we expect to 
see ultraluminous infrared galaxies (ULIRGs) with peak IR luminosities $L_{IR}$
in the range of $10^{12}$ - $10^{13} L_{\odot}$ based on the local UILRGs
observations, though lower luminosity-ranged mergers had been observed
as well presumably between smaller galaxies or less gas-rich systems 
(Sanders \& Mirabel 1996, Table 3).  
The assumptions we made to arrive at the numbers used in Table 1 for mergers are 
that the galaxies involved in these mergers are {\em large}, as well as {\em gas-rich} 
disks, a condition clearly satisfied by the BO-effect progenitor galaxies 
(Dressler et at. 1997).  For the virializing region of a dense cluster,
if we expect major mergers to be responsible for producing the observed level
of BO effect, or the transformation of F = 30\% of the galaxy population 
from spirals to S0s and other early types in the past 5 Gyr,
and if we assume an interaction timescale of $5 \times 10^{7} - 1.5 \times 10^{8}$ yr for each of
the two higher luminosity ranges listed in Table 1 (Sanders \& Mirabel 1996, who further
cited the merger simulations of Mihos \& Hernquist 1994; also Murphy et al. 1996),
equation \ref{eq:eqfraction} gives an observed merger fraction of 
$f_{merger}$ = 0.3 - 0.9\% for each of the higher IR luminosity ranges. 
We have also assumed the BIRGs (bright IR galaxies, with $L_{IR}$ in 
the range of $6 \times 10^{10} L_{\odot} - 4 \times 10^{11} L_{\odot}$) phase to last
about twice as long as the average  
VBIRGs (very bright IR galaxies, with $L_{IR}$ in the range of $4 \times 
10^{11} L_{\odot} - 10^{12} L_{\odot}$) and ULIRGs phases, i.e., it has a timescale
of $2 \times 10^8 yr$, tracing 
precursors and aftermaths of merger events, which leads to the estimate 
of $f_{merger}$ = 1.2\% in the BIRG phase. 

We note that different individual merger
events have different interaction kinematic configurations, which will
also result in the difference in how long the merging process will last.
This uncertainty in the timescales of merging events should not hamper our
ability to determine the possible roles of merger in a given cluster,
since the predictions of the merger scenario, both in terms of observed
fractions and in terms of the luminosity of its most luminous members, are orders of
magnitude different from other scenarios, and thus can tolerate this level
of uncertainty.  In the example we present in the next section, we will see that 
the predominence of moderately-luminous IR sources
in that cluster from many different angles precludes the possibility
that merger had played a significant role in transforming the galaxy morphologies
in that cluster.

For the outskirts of the cluster we assumed have a morphological transformation rate of $F=0.1$.  
This leads to an average $f_{merger}$ = 0.2\% expected for the outskirts if merger is
the major cause of the BO transformation, slightly higher than for the field 
of $f_{merger}$ = 0.1\% which is the observed value (Sanders \& Mirabel 1996). 

\subsection{Galaxy Harassment}

Since galaxy harassment is a mechanism proposed based on numerical
simulations (Moore et al. 1996,1998), rather than on observations
of physical processes, the precise timescales of this mechanism
are not prescribed.  We expect, from the general dynamics of 
high-speed swing-by between galaxies, that the timescale
should be on the order of $10^8$ yr for each of the BIRG and VBIRG
phases, similar to the less-violent phases of a major merger.  
Based on this assumption, as well as the proposed
{\em multiple} nature of the harassing encounters, which we take to be 5 for
an average galaxy to go through the late-to-early-type transition
through the harassment process, we arrive at an overall time
scale $t_{event}$ of $10^9$ years (5 times of $10^8$ years in each of the
two luminosity ranges), which is reasonable considering
that this is essentially the timescale of the cluster-core crossing
for cluster galaxies.  The high-density cluster core region is
believed to be where the harassment mechanism is most effective
(Moore 2004). These timescales gave the estimates of the 
observing frequencies quoted in Table 1.

The harassment process is expected to be much less effective
away from the high-density cluster core regions, since it depends on the
strong tidal interaction between close-neighbor galaxies.  We therefore
set the expected observing frequency for IR-bright galaxies due
to harassment in the outskirts of clusters to be nearly zero.  Any
observed IR bright galaxies in those regions must be due to
physical processes other than galaxy harassment.
We would not expect to see ULIRGs either in the core or in the
outskirts due to the action of galaxy harassment process alone.

\subsection{Ram Pressure Stripping}

Like the harassment mechanism, the ram pressure stripping mechanism
is also expected to operate mostly in the dense region of a cluster,
since that is where the intracluster medium (ICM) is mostly concentrated.
If ram pressure is the dominant mechanism for the BO transformation, 
we would expect BIRGs to be preferentially located in the central 
virializing region of clusters, since this is a gentle process which is not expected
to produce ultra-bright IR sources.  Furthermore, the timescale
of operation of this mechanism is on the order of a cluster-core
crossing time, or about $10^{9}$ yr (Bosseli \&
Gavazzi 2006), though shorter timescales (down to $10^8$ years)
have been proposed as well (Quilis et al. 2000;
Fujita \& Nagashima 1999).  This leads to the frequencies
of observations derived for the active virialization region and the
outskirts of a cluster due to
the operation of ram-pressure stripping alone, as shown in Table 1.
We would not expect to see ULIRGs in any regions of the cluster
due to the ram pressure stripping process alone.

\subsection{Post-Interaction Secular Evolution}

The post-interaction secular evolution process refers to the enhanced
star-formation and radial mass accretion process mediated by the
large-amplitude, open density wave patterns such as spirals, bars,
as well as skewed three-dimensional mass distributions (Zhang 1999;
Zhang \& Buta 2007a,b).
Secular evolution can operate in both the core as well
as the outskirts of a cluster (or even in the
group environment or in isolation), the only difference between
these environments is the
different evolution rates caused by the different strengths of the
density waves, which in turn are caused by the different amount of
tidal perturbation in these environments. 

For the tidally-enhanced secular evolution process we expect a large 
population of BIRGs with luminosities up to the lower range of VBIRGs.  
The IR manifestations of the secular evolution process
depend not only on the strength of tidal perturbation, but
also on the availability of gas supplies for star-formation.  
In the post-virialization region, even though the mass accretion process
may still be going on, the star formation activity can be significantly
quenched due to the paucity of gas in this region of the galaxy disk.

For the active virializing region, since the member galaxies are effectively
always feeling the gentle tidal nudges of the near and as well as far
neighboring galaxies (a condition characterizing the virialization
itself), we therefore set the highest timescale of secular evolution in the
BIRGs phase to be $5 \times 10^{9}$ yr, i.e., exactly equal to
the amount of elapsed time between z=0.4 and z=0.  This of course
assumes the continuous availability of gas during this time period.
To take into account of the more rapid consumption of gas (i.e., a shorter
star-formation timescale than the morphological-transformation timescale),
we set the lower limit to be half of the maximum. Note that these timescale
estimates for secular evolution assumes spiral-induced star formation and mass accretion
which are a disk-specific processes.  There could exist
more violent primordial, clumpy mass accretion at higher z due to dynamical
friction in proto-galaxies, as might be what have terminated the
star formation in large and early-virialized clusters such as MS0451,
which we will discuss later.

For the outskirts we have assumed the star formation timescale to be about $1.5 \times 10^9$ yr,
typical of post-starburst evolution timescale (Poggianti et al. 1999),
since the interaction frequency at the outskirts is expected to be low,
and interaction is not continuous as in the active virialization region.

With these assumptions we derive the expected fractions of the BIRG sources
as given in Table 1.  The fraction of the VBIRGs are calculated
assuming an interaction timescale of $10^8$ yr, describing the precursor 
starburst events leading to {\em some} of the subsequent secular evolution.
We would not expect to see ULIRGs due to the secular evolution process
alone.

\bigskip
\bigskip

\begin{table}[ht!]
\begin{tabular}{|c|c|c|c|c|} \hline
&$f_{merger}$ & $f_{harassment}$& $f_{gas~stripping}$&$f_{secular ~ evolution}$ \\
\hline
 post-virialization& N/A& N/A& N/A& N/A\\
region& & & & \\
\hline
& $\sim$ 1.2\% BIRGs & $\sim$ 3\% BIRGs& $\sim$ 0.6\%-6\% BIRGs & $\sim$ 15-30\% BIRGs\\
& (12) & (30) & (6-60) & (150-300) \\
active & $\sim$ 0.3-0.9\% VBIRGs & $\sim$ 3\% VBIRGs& $\sim$ 0\% VBIRGs & $\sim$ 0.6\% VBIRGs\\
virializing & (3-9) & (30) & (0) & (6) \\
region& $\sim$ 0.3-0.9 \% ULIRGs & $\sim$ 0\% ULIRGs& $\sim$ 0\% ULIRGs & $\sim$ 0\% ULIRGs\\
& (3-9) & (0) & (0) & (0) \\
\hline
& $\sim$ 0.4\% BIRGs & $\sim$ 0\% BIRGs& $\sim$ 0\% BIRGs & $\sim$ 3\% BIRGs\\
& (4) & (0) & (0) & (30) 
\\ outskirts & $\sim$ 0.1-0.3\% VBIRGs & $\sim$ 0\% VBIRGs&  $\sim$ 0\% VBIRGs & $\sim$ 0.2\% VBIRGs\\
& (1-3) & (0) &  (0) & (2) \\
& $\sim$ 0.1-0.3\% ULIRGs & $\sim$ 0\% ULIRGs& $\sim$ 0\% ULIRGs & $\sim$ 0\% ULIRGs\\
& (1-3) & (0) & (0) & (0) \\
\hline
\end{tabular}
\bigskip
\caption{The expected fractions and numbers of galaxies to be observed in each
luminosity range for z $\sim $ 0.4 rich clusters in two different cluster
environments, for different proposed physical processes acting alone, generated
using equation (2).
The expected number of detected galaxies for the different
scenarios are given in parentheses, assuming 10 clusters, each having
an average of 200 galaxies in the two outer regions (and therefore 
approximately 100 galaxies in each of the two regions described above). 
Here ULIRGs (UltraLuminous
InfraRed Galaxies) have IR luminosities in the range of $L_{IR}
= 10^{12}$ - $10^{13} L_{\odot}$, (Sanders \& Mirabel 1996),
VBIRGs (Very Bright InfraRed Galaxies) have $L_{IR} = 4 \times 10^{11}$
- $10^{12} L_{\odot}$, and BIRGs (Bright InfraRed Galaxies) have
$L_{IR} = 6 \times 10^{10}$ - $4 \times 10^{11} L_{\odot}$.
}
\end{table}

Finally we want to comment here that in this entire derivation for
the frequency table (Table 1), we have implicitly assumed: (1) the constancy
of the rates of the interaction processes across the redshift range
under concern, (2) the availability of the gas supply to feed the
star-formation activity at the epoch of the observation z, and
(3) a single process dominates the BO transformation.  All
of these assumptions may need to be modified, at least for some
clusters and during some periods of the evolution.  So corroborating
data from other wavelengths for the clusters under study will help
to arrive at a satisfactory picture of the major drivers for the morphological
evolution in a given cluster, or in a population of clusters
studied in a statistical sense.  Apart from these implicit assumptions
the only uncertainty of the Table quantities comes from the
timescale value $\tau_{event}$ for each proposed physical process
(since $F$ and $T$ in equation 2 are well-determined
quantities and are common for all processes in a given cluster).

\section{An Example of the Application of the Proposed Approach}

The recent 24 micron Spitzer observations of two distant clusters 
(Geach et al. 2006) have revealed elevated levels of star formation 
throughout the cluster up-to and slightly-beyond the cluster 
turn-around radius, both compared with that determined 
using optical observations and compared to values typical of the field 
environment at the same redshift.  It also revealed very different frequencies 
of IR-luminous sources in these two clusters: For MS 0451-03 at z=0.55,
very few 24 $\mu m$ sources were detected; whereas for CL 0024+16
at z=0.39, a large excess of the 24 $\mu m$ sources were detected
($\sim$ 150 IR galaxies over a 25'x25' area, or 9x9 $Mpc^2$).  

The difference in the rate of IR luminous sources may reflect a correlation of the IR sources 
with the virialization state of the cluster, i.e., even though
MS 0451-03 is at slightly higher z than CL 0024+16, it has larger mass
and also appears to be at a more advanced stage of
virialization than CL 0024+16 (Geach et al. 2006; Moran et al. 2007b), and therefore
the most intense star-formation episodes may have already been over
by the observation epoch (as evidenced by the presense of many passive
spiral galaxies in this cluster, Moran et al. 2007b).  In this case the 
entire cluster belongs to region 1 (post-virialization region) of the Table, 
which can be consistent with essentially all of the proposed mechanisms (i.e.,
we will not be able to discriminate among the proposed mechanisms
for this particular cluster).  To sum up, the lack of IR sources in
MS 0451 may indicate the operations of early-epoch physical processes 
(such as clumpy mass accretion due to dynamical friction or spherical mass accretion)
prior to the observational epoch which had effectively
suppressed star formation before any other processes
(such as secular evolution) would have an opportunity to
generate an IR-bright starburst at the observational epoch, though
the mass accretion process due to stellar density wave structures
in the passive spirals in this cluster could still
continue to modify galaxy morphology without generating the
corresponding starburst activity.  The star-formation
condition of MS 0451 is not expected to be typical for an average
BO cluster at intermediate redshifts, though, since BO clusters
are usually seleted to contain a large population of gas-rich star-forming
galaxies.  We will focus our attention of the application of the proposed approach
thus only on CL 0024+16.  

It is well known that in disk galaxies star formation happens
mostly around the spiral arms or bars, i.e., around the density
wave crest.  This is because the disk is usually stable to the
gravitational collapse of matter, and a density wave shock (which
is a collisionless shock induced by the nonlocal gravitational
potential wave which is phase-shifted with respect to the density distribution)
is needed to trigger a violent episode of star formation (Zhang 1996 and the references therein).
The reason that interactions usually enhance the amplitude of the
spiral density wave is because the spontaneous growth rate of the
density wave is usually very low (Zhang 1998), so many field spirals
can have very low amplitude density waves, and remain in the disk
phase even until the recent time, whereas in the cluster environment,
the tidal interactions remove angular momentum from the disk galaxies,
and also excite large seeds of non-axisymmetric perturbations which are
further amplified by the disk's intrinsic modal resonance mechanism.

Most of the 24 $\mu m$ galaxies in CL 0024+16
observed by Geach et al. (2006) show enhanced luminosity in the range of BIRGs, or
$6 \times 10^{10} L_{\odot}$ to $ 4 \times 10^{11} L_{\odot}$,
compared to their counterparts in the field ($ \sim 2-3 
\times 10^{10}$ $L_{\odot}$ for normal disk galaxies, see Sanders \& Mirabel 1996).  
The majority of the 24 $\mu m$ sources detected by Geach et al. (2006)
displays an optical/NIR colors that are intermediate between star-burst galaxies
and passive galaxies, which indicates that this is likely a population of 
galaxies in the post-interaction stage, and are going through the slow
and prolonged evolution under the influence of the interaction-enhanced density-waves.  
We now apply the frequency approach developed in the last section to further
confirm the nature of these 24 $\mu m$ sources.

Geach et al. (2006) found that there are
about 150 IR luminous sources in an area of 25'x25', 
the majority of them are in the IR luminosity range
of $ 6 \times 10^{10} - 2 \times 10^{11} L_{\odot}$, which,
compared with the total galaxy number of about 500 determined
by Moran et al. (2005) for the same region, 
and taking into account of completeness
estimates ($\sim$ 65\% completeness for galaxies in the 
range of $17.75 < I < 21.1$ and $\sim$ 40\% for galaxies in the range 
of $17.75 < I < 22.5$), leads to an estimation of
a fraction of about 15\% of IR-bright sources among
the entire optical sample\footnote{In arriving at these
fractional estimates we have considered the fact that we might need to
extend the lower bound of the IR flux range from
$6 \times 10^{10} L_{\odot}$, which is the 5 $\sigma$ 
detection limit of the Geach et al. observations,
to closer to $3 \times 10^{10} L_{\odot}$ which is the typical flux level of 
the local dusty starburst galaxy, such as NGC 253.  However, we note
that not all dusty starburst at the lower IR luminoisty range will
lead to the conversion of a late-type to an early-type galaxy in 5 Gyr,
therefore the more numerous IR sources at the lower luminosity range
probably should not enter into the current statistics. This consideration
is also supported by the recent result of Elbaz et al. (2007) who found
higher redshift starburst galaxies to be much more IR luminous on average
than the local starburst galaxies of similar morphology.}.
Note that Geach et al. (2006) did not carry out the luminosity
statistucs of their 24 $\mu m$ sources according to the different
radial ranges in the cluster, but rather estimated the luminosity
statistics for the cluster as a whole.  Therefore we do not
have more detailed information at the present time to do a more
detailed analysis in terms of the radial distributions of
the frequency of the different kinds of IR bright sources.

The derived fraction of IR-bright sources
is in the correct range as predicted by Table 1 for
secular evolution to account for the majority of the BO transformation
between the intermediate redshift and the present for this cluster:
the fraction predicted in Table 1, when averaged over the entire cluster
region, would lead to a predicted BIRG fraction of around 15\%.
From Geach et al. (2006)'s result, it is also seen that the
brighter IR sources (those in the VBIRG range of $4 \times 10^{11}
- 6 \times 10^{11} L_{\odot}$) consist of about 4 sources, or 0.4\%,
which is also consistent with the secular evolution scenario.
Most significantly, there were no sources
detected above the flux level of $6 \times 10^{11} L_{\odot}$, or no
ULIRGs or strong interactions present in this cluster
at this observing epoch.  
Improved statistics from a much larger sample would help to quantitatively assess the roles
of other proposed mechanisms across the entire population of
BO clusters.

Dressler et al. (1999) showed that the post-starburst galaxy
fraction in their sample of 10 intermediate redshift clusters
observed using the HST in the MORPHS project is about 20\%, similar
to the fraction of IR-bright sources in CL 0024+16. 
Poggianti et al. (1999) confirmed that the total
a+k/k+a type galaxy(the type of emission-line
galaxies signifying a post-starburst population) fraction is
about 20\%.  They had had considerable trouble identifying
the progenitor population of the post-starburst galaxies,
however, presumably because these last only a brief period
and therefore are less numerous.  

In the MORPHS sample analyzed by Poggianti et al. (1999) 
and Dressler et al. (1999), 10\% of cluster galaxies were clasified
as e(a) (which is a dusty-starburst spectral class), and about
40\% of these (or 4\% of the total) showed morphological signs of
tidal interaction or were observed to be directly involved in a
merger/close interaction. This is much larger that the maximum
allowed fraction of mergers we had estimated in Table 1 (i.e., 0.9\%),
which was calculated based on the observed BO transformation rate
between z=0.4 and the local universe.
We therefore felt that what Poggianti et al. and Dressler et al. observed 
could be not all mergers but rather interactions that were confused
with mergers due to limited spatial resolution.  This conclusion is
also supported by the fact that in both CL 0024 and MS 0451, the Spitzer
24 um MIR observations by Geach et al. (2006) had revealed essentially
no ULIRGs population at all. 

There exists the question of whether the few highest luminosity sources (currently
estimated to be in the mid range of $10^{11} L_{\odot}$ in
the Geach et al. observations) in CL 0024
could have been undergoing mergers, because there is uncertainty in
the conversion of 24 $\mu m$, or the rest frame $15 \mu m$ flux to the IR luminosity.  We argue that this
possibility could not be true for this cluster, based on the entire distribution
of the IR sources in this cluster. In Table 1 the predicted rates for mergers are
not only for the ULIRGs, but also for the BIRGs and VBIRGs.
For a given dynamical scenario the IR source statistics in
all three ranges of the IR luminosity need to be
accounted for, not just the highest luminosity range.  The over-abundance of 
moderately IR-bright sources in this cluster is not
consistent with a scenario in which merger is the dominant process for BO transformation
in this cluster, because if so a roughly 0.3-0.9\% ULIRG fraction would
by itself be able to account for the F=30\% BO
conversion, which leave the rest of the 15\% moderately luminous IR sources unaccounted for.
Or if these moderately bright IR source also lead to transformation by some other dynamical
mechanisms such as secular evolution in addition to the role played by merger, together they
would be over-producing the observed 30\% BO conversion ratio between z=0.4 and present.

Furthermore, the observed IR sources could also not have been consistent with their
being the product of either harassment or ram pressure stripping,
since these should have produced fewer IR sources based on the
values given in Table 1.  Plus, harassment should have erased
the spiral structures whereas most of these post-interacting galaxies
were found to be spiral disk galaxies (Poggianti et al. 1999; Moran et al. 2007b). 
Ram pressure stripping would also have difficulty producing the
brighter portion of these IR sources in CL 0024.  Thesefore, tidally-enhanced
secular evolution process becomes the only possible remaining candidate
to account for the origin of the population of IR sources in CL 0024.

Other recent ground-based optical observations on CL 0024 using other instruments
appear to corroborate our conclusions about the role of interaction-enhanced
secular evolution in accounting for these excess IR sources. Moran et al. (2007a)
found from the analysis of resolved optical spectroscopy of CL 0024
galaxies and a control sample of field galaxies at similar redshifts that the
cluster Tully-Fisher relation exhibits higher scatter than its field
counterpart.  They also found that the central mass densities of
the spiral galaxy population they examined were higher within the
cluster virial radius than outside, with a sharp break exactly
at the cluster virial radius.  The cluster environment thus appears to be
responsible for the creation of the increase in scatter of the Tully-Fisher
relation, as well as for the increase in central density of cluster
spirals within the cluster virial radius.

Moran et al. (2007a)'s tentative explanation of these observations
is that galaxy harassment process has eliminated the less-dense
spirals in the cluster central regions.  This explanation however
could not account for the lack of high-central-density spirals in
the cluster outskirts.  So they proposed that these cluster-outskirts
high-density spirals might have been eliminated by mergers.
However, an obvious problem with this proposed explanation is why
mergers in cluster outskirts would selectively eliminate only
the high-density spirals but not the low-density spirals.
Furthermore, 24 $\mu$m observations of this cluster have not revealed
a significant merging population in this cluster.

The observations of Moran et al. (2007a) were also difficult to
explain through any kind of ram-pressure striping mechanisms,
as these usually lead to unusual star formation gradient,
whereas the enhanced scatter observed in Tully-Fisher relation
was present in both the V and the $K_s$ band, and thus is not due to
the influence of star-formation or dust obscuring, but rather
has to be related to the kinematic and structural changes of
the cluster galaxies, as Moran et al. (2007) had concluded.

The observed trend would be most consistent with a mild level
of tidal perturbation for cluster galaxies (and these perturbations
are obviously more pronounced for galaxies within virializing region
of the cluster, because the condition of virialization
essentially guarantees a continuous interacting state of the
member galaxies in the cluster virial radius), as well as the
post-interaction secular evolution of the galaxy morphology and 
kinematics.  

The secular evolution scenario could naturally produce the 
observed difference in the central mass density of disk galaxies 
inside and outside the cluster virial radius, as well as the increase 
in structural and kinematic scatter of these disks (Moran et al. 2007a).
Note however that the increase
of scatter is only observed for the Tully-Fisher relation
of the late-type cluster galaxies within the cluster virial radius.
For S0s, for example, the fundamental scaling relation in fact
becomes tighter within the virial radius (Illingworth et al. 2000).
This phenomenon is easily explained in the secular evolution
scenario as the settling onto an ``attractor'' of the dynamical
evolution (see \S 5).

Furthermore, the fact that the blue members of a cluster
show more scatter when placed on a scaling relation plot, whereas
the red ones show better correlation, shows that
the interactions in the cluster environment tends to disturb the
perfect Tully-Fisher relation for late-type cluster galaxies compared
to the field.  However, as the secular evolution proceeds, the 
cluster environment in fact produces the tightest scaling relations 
at the late stages of evolution, much tighter than the counterpart
of these scaling relations for the field early-type galaxies.
The tidal agitation in the cluster environment is a way to disturb
one dynamical equilibrium (that of late type disks)
to facilitate the settling onto a new dynamical equilibrium
(that of early-type morphology, which is of higher gravitational
entropy).  

Once again, we emphasize that the conclusions we have reached based
on the analysis of CL 0024+16 serves only to illustrate the application
of the approach we proposed, and the conclusion on the dominating
mechanism to account for the BO effect for the majority of
the intermediate redshift clusters can only be reached once we have 
analyzed enough samples of rich clusters by the same approach.
At the time of this writing, CL 0024 and MS 0451 are the only 
intermediate-redshift clusters 
that have been observed in this fully-sampled wide-field fashion among
the Spitzer-approved observations. A few more will be observed by
Spitzer in Cycle 4.  Numerous other observations
do exist, however, for the central, virialized cores of clusters,
and can be used to further carry out the existing program.  As
we have seen in this preliminary study, even the information
obtained in the core region already gave plenty of hints at the
underlying dynamical mechanism responsible for cluster galaxy
transformation, though the addition of the cluster ourskirts
and group observations will help to complete the picture.

\section{Other Corroborating Evidence in Support of a Secular
Evolution Origin for the Morphological Butcher-Oemler Effect}

\subsection{Results from Sloan Digital Sky Survey}

Goto (2005) analyzed the
velocity dispersion of 355 galaxy clusters (with 14548 member
galaxies) from the Slone Digital Sky Servey (SDSS), and found
that bright cluster galaxies have significantly smaller
velocity dispersion than faint galaxies, consistent with
a picture of dynamical friction (galaxy-galaxy interaction)
operating during the process of cluster virialization which 
reduces the velocity dispersion of massive cluster galaxies at the 
expense of the increase of velocity dispersion of the less massive 
cluster galaxies.  He also found that star-forming late-type 
galaxies in his sample have a larger velocity dispersion than the
passive late-types (i.e. those having spiral morphology,
but do not show any ongoing star-formation activity, see 
Goto et al. 2003b), which is once again consistent with dynamical friction
reducing the velocity dispersion of the more evolved (passive)
population of galaxies.  

The result of Goto (2005) is consistent with our
conclusion reached from CL 0024+16 that tidally-enhanced
secular evolution appears to be the driver for
transforming the cluster galaxy morphology, since the galaxy-galaxy
interaction and dynamical friction is known to excite large-amplitude
density waves which leads to subsequence radial accretion of
mass and the evolution of galaxy morphological types (Zhang \&
Buta 2006). The Geach et al. (2006) and Goto et al. (2003b), Goto (2005)
results are also mutually consistent in the sense that the 24 $\mu$m
sources which show intermediate colors between star-burst and
passive populations are likely to be the population in transition
from star-burst to passive galaxies, and these passive galaxies,
through further secular evolution, will become the earlier Hubble-type
cluster galaxies observed in nearby clusters.

The role of dynamical friction during the cluster relaxation process
is also reflected in the often sharp transition in galaxy
properties across the cluster virial radius, or at a critical
local surface density (Moran et al. 2005,2007; Tanaka et al. 2004;
Couch, Colless, \& Propris 2004. Figure 5 of Illingworth et al. 
2000 shows another example of the change of the tightness of
scaling relation for the S0 population across the cluster
virial radii). These characteristics, as we commented before, can be
naturally explained by the enhanced secular evolution due to
the tidally-induced interaction processes.

Goto (2005)'s result is on the other hand inconsistent with
ram pressure having played a significant role in causing
cluster galaxy evolution. The effectiveness of the ram pressure is proportional
to $\sigma v^2$ (where $\sigma$ is the gas density and $v$ is the velocity
dispersion), which should be more effective for galaxies with
larger velocity dispersion, i.e., it should have predicted that
the larger-velocity-dispersion population is the
more evolved one (i.e. passive), contrary to what is observed.

\subsection{Results from the MORPHS Collaboration}

The MORPHS team conducted imaging and spectroscopic studies
of 10 distant clusters in the redshift range of 0.37 to 0.56.
Poggianti et al. (1999) found that among the MORPHS cluster
galaxies a significant over-abundance of the
 so-called ``post starburst" galaxies with the
characteristic k+a/a+k spectral features, which exhibit spatial
and kinematic distributions intermediate between the passive and
active populations. The number density of these post-starburst
population is significantly higher in their sample clusters than 
in the field, so there the star-burst activities prior to the k+a/a+k 
phases were obviously triggered by the cluster environment, so are 
the subsequent termination of the star-forming events.

One interesting thing noticed by Poggianti et al. is the mostly
spiral-like disky morphology of these post-starburst galaxies from
the HST images, showing that whatever interaction processes
which triggered the starburst had not changed the galaxy
morphology immediately. This fact is contrary to both the merger
and the harassment predictions, which require immediate change
to galaxy morphologies after the interaction.  The evidence for the
milder tidal perturnation, on the other hand, is entirely
consistent with an interaction-enhanced secular evolution scenario.

Poggianti et al. also concluded that either two timescales
or two different physical processes has caused the two observed
transformation processes (the halting of star formation and the spiral-to-S0
transformation).  This result is supported by the later investigation
of Goto et al. (2004) using the SDSS data, who found that the color
evolution of the BO galaxies in their sample is much faster than the
morphologial evolution. In the secular evolution scenario, 
these two different timescales acquire a most natural explanation:
the interaction events which serve to excite the large-amplitude
spiral density waves and bars may only last on the order of
0.1-1 Gyr, but the subsequent secular mass accretion process
could continue for much longer, i.e., could be a significant fraction of a Hubble
time.  Furthermore, the same process could operate on the passive
phase of the galaxies involved as well, even when most of the
star-formation activities have extinguished.

\section{Further Comments on the Different Proposed Mechanisms
for Cluster Galaxy Evolution}

\subsection{Hierarchical Clustering and Mergers}

The Hierarchical clustering/CDM paradigm has been our standard
paradigm for structure formation over the past few decades.
However, growing evidence has pointed to the inadequacies of
using this paradigm to explain the observations especially
on the properties of individual galaxies.

The well known down-sizing trend of galaxy formation, not only
in terms of star formation history but also in terms
of mass assembly history (Cowie et al. 
1996; Kodama et al. 2004, Cimatti, Daddi \& Renzini 
2006 and the references therein)
is the most obvious contradictory evidence to the bottom-up
assembly scenario which is the foundation of the CDM paradigm.  
While the former trend (down-sizing in star-formation)
can perhaps be accounted for by proposed mechanisms such
as AGN or star-formation feedback (see, e.g., de Lucia et al.
2006; Croton et al. 2006 and the references therein), the
latter trend of mass assembly down-sizing (Cimatti et al. 2006)
is in direct conflict with the mass assembly history
predicted by the CDM paradigm.  The observed
early formation age of cluster early type galaxies (Franx 2004
and the references therein), even
after taking into account some late formations
due to the BO effect, is also contrary to the
predictions of the CDM paradigm.  CDM is known to significantly
under-predict the extremely-red-objects (ERO) observed in the early
universe (Daddi, Cimatti, \& Renzini 2000 and
the references therein).  It also predicted significant difference in the properties
of the field and cluster early-type populations (Kauffmann 1996), which is
not observed.  Other evidence for the difficulty of using 
this paradigm to explain the observed properties of galaxies 
can be found in Zhang (2003).

CDM predicted the late assembly of early type galaxies,
including cluster galaxies, which superficially is consistent
with the morphological BO effect and with
the increase in number density of early type galaxies
observed for both the cluster and field (Treu 2004 and the references therein).
However, unequal-mass mergers completely destroy the
color-magnitude relation (Bower, Kodama \& Terlevich 1998),
whereas equal-mass merging among massive galaxies
reduces the early-type number density, contrary to the
observed early-type number density evolution.
Furthermore, collisionless mergers simply can't reproduced
the high phase space density observed for the
central regions of early-type galaxies (Ostriker 1980).
The rotationally-supported S0 disk, which is the end product
of the majority of BO transformation in clusters, is also problematic
for a merger-based morphological transformation process.
These observational constraints, coupled with the high velocity dispersion of
cluster galaxies, means that mergers, including dry mergers,
could not have played an important role in the BO
transformation, a conclusion consistent with our finding
in this work based on the analysis of the properties
and frequencies of the IR sources for CL 0024.
The recent analysis by Elbaz et al. (2007) on the GOODS fields
data also directly confirmed that a great portion of the
star-formation events at $z \sim 1$ is confined within single
disk galaxies, rather than due to the merging of different galaxies.
The star-formation rates observed in the GOODS fields also
have drastically different dependence on the densities of galaxies
than that predicted by the CDM paradigm (Elbaz et al. 2007, Fig. 8). 
The observed density dependence of star-formation rate, however, is
consistent with the expectation from the interaction-enhanced
secular evolution scenario.

\subsection{Infall, Stripping, Mixing, and More on the CDM Paradigm}

Poggianti et al. (1999) found no evidence among the MORPHS
cluster galaxies of a difference in the radial distributions
of passive and emission-line spiral galaxies, though the recent result
of Moran et al. (2007b) is somewhat in conflict with it.
A difference of distribution in the distribution of
these two populations is to be expected
if significant infall had occurred.
Cooper et al. (2006,2007) described the results from the DEEP2
survey, and found that the steepening of the morphology-density 
relation works in the group environment
just as effectively as in the cluster environment.
The results from SDSS and 2dFGRS also support this conclusion.
Therefore, cluster-specific mechanism, such as ram-pressume stripping, 
will not be a justified cause for the group galaxy evolution.
Dressler et al. (1997) concluded that the morphology-density 
relation indicated that more local processes are at work.
Couch, Matthew and De Propris (2004) reached 
similar conclusion from the 2dF survey results.

The new results of Elbaz et al. (2007) showed that at the
high redshift the mechanism which relates galaxies to their
environment is not simply a quencher but also a trigger
of star formation: the star-formation/density relation at z=1
in fact reverses the morphology-density relation trend.
These authors concluded that their results are most consistent
with a scenario where star formation is accelerated by the
dense environment of galaxies, resulting in the faster exhaustion of
their gas reservoir.  This new result of the star-formation
and galaxy environment dependence is inconsistent with the
ram-pressure stripping picture. 

As we have commented earlier, the existence of the
morphology-density relation (Dressler 1980; Dressler et al. 1997)
means that there could not have been a significant amount of 
persistent mixing of late type galaxies from the field environment into
the cluster environment during the cluster evolution process.
Similar minimal-mixing conclusions have also been reached by
Abraham et al. (1996) and Morris et al. (1998). The hierarchical
assembly scenario, on the other hand, advocates the gradual assembly of higher
density clusters from lower density sub-clusters, which is slightly
different from the continuous infall scenario for individual
galaxies.

Through numerical simulations, Dressler (2004) and collaborators
found that under the CDM paradigm there is actually significant mixing 
on cluster dynamical timescales, at least on the group level.
This predicted mixing by the CDM, however, is in conflict with
the observational evidences quoted above, including the existence 
of the morphology-density relation.  This in fact from another 
side highlights the inadequacy of one of the
central assumptions of the CDM paradigm (see, e.g., Peebles 1993): 
that of cluster formation as a purely
gravitational process.  This assumption is also what went into all
the cluster CDM simulations which resulted in the observed mixing of
the galaxy populations in the different regions of a cluster,
contrary to the observational evidence of
well-segregated evolution manifested in the morphology-density relation.  

There are in fact very well established observational evidence that
the cluster formation process is not a purely gravitational process.
Evrard (2004) reviewed many existing theoretical and observational
tests of cosmology using galaxy clusters as probes, and highlighted
the deficiency of the cluster physics according to the CDM paradigm
as ``a cluster energetics problem''.  Under the assumption that the
principle mechanism of ICM heating is through gravitationally induced
shocks, there is about a 70\% excess heat that could not be accounted
for between what is required for virial equilibrium and what is observed.
Furthermore, galaxy velocity dispersion in clusters shows a comparable
level of excess ``heat'' compared to the dark matter velocity dispersion
normalized to the WMAP and large scale structure distributions.
Therefore, the cluster environment is in general not in a gravitationally 
unstable collapsing configuration as predicted by the CDM theories.
Depending on gravitational collapse alone
most of the clusters as we know would not have formed.
Many of the observed clusters form through their inherent hydrodynamical
large-scale systematic velocities and the resultant high-speed sub-cluster clump
mergers (such as what had happened for CL 0024, as well as for many other
well-known clusters in the intermediate redshifts as studied by the MORPHS collaboration
and by Couch et al.), but the cause of these mergers are not gravitational
collapse, but rather hydrodynamical collision due to their inherent relative
velocities. A portion of the observed clusters such as MS 0451 do indeed
appear to form out of gravitational collapse, however. 
These clusters are likely to be situated in a high
density region in the primordial mass fluctuation spectrum,
and thus are much better virialized as a whole because they satisfy better
the gravitational instability condition. 
This latter class will appear to be more X-ray
luminous for a given amount of baryonic mass. They also tend to evolve faster
in terms of member galaxy morphologies as the example of MS 0451 had shown. 

The excess systematic velocities of many of the
proto-cluster clumps were thought to have originated from the so-called ``primordial
turbulence'' at the time of CMB decoupling, an idea which can be traced back to the early
proposals of von Weisacker (1951) and Gamow (1952), and which was subsequently
substantiated by Ozernoy (1974a,b, 1978).  This idea has
recently been shown to be relevant also to the
explanation of the angular spectrum of the the observed cosmic microwave background
radiation (Bershadskii \& Screenivasan 2002,2003).
Peebles (1974, Figure 1) had plotted the covariance function
for the angular distribution of galaxies, and found that a power law fluctuation
index n between -1 and 1 can fit the observational angular distribution of galaxies.
A power law index of n=0.33 can originate from the Komogorov type of turbulence
velocity spectrum, and the dissipation in turbulence can steepen the
residual velocity spectrum above 1/3, to close to the WMAP observed
value of n $\approx$ 1.
If the primordial turbulence underlies the formation of the large scale
structure and the clusters, then the paradoxial excess heat of the
ICM and excess velocity dispersion of the cluster galaxies
(Evrard 2004 and the references therein) would all acquire a very
natural explanation, which the current CDM paradigm and the gravitational
collapse picture failed to provide an explanation.
In the primordial turbulence picture mergers due to inherent systematic
velocities between proto-cluster clumps will play a much more important
role, and infall and mixing which are associated with the CDM gravitational
collapse will not be an important physical process during the
cluster formation and evolution process.

Even though the CDM paradigm has been successful in explaining
many features of the large-scale structure observations (though
by no means this is the only possible way to explain it, as our above
references had shown), the observations on galaxy-scaled phenomena 
is the area that the CDM paradigm has run into the most serious problems,
both in terms of the mass profiles
of galaxies along the Hubble sequence, for which it predicted the
progressively-later Hubble types to have a more cuspy mass density,
exactly opposite to what was observed, which shows that it is
the early types that possess the cuspy surface density; as well as in terms of the
mass assembly history, which is currently found to be anti-hierarchical
(Cimatti et al. 2006).  Secuclar evolution
cannot be simply inserted into the CDM paradigm to help solve its many problems on
the galaxy level, as many had hoped,
because as an initial condition secular evolution assumes a physical low surface brightness (LSB) 
galaxy surface density profile, which
is flat across the disk, not the cuspy type predicted by the CDM as a
result of cold dark matter quickly sinking into the center of a galaxy.

\subsection{Galaxy Harassment}

Countrary to the later usage by many investigators, the early proposers
of harassment intended it mainly as a mechanism for the transformation
of dwarf disk galaxies to dwarf spheroidal galaxies (Moore et al.
1996).  The harassment mechanism has been shown to be
ineffective during the simulations for transforming the morphologies
of large disks, especially in increasing the bulge-to-disk ratio (Gnedin 1999;
Moore et al. 1999). As is well known, a large
fraction of the disk galaxies undergoing BO transformation are
normal spiral galaxies which have large disks (Dressler et al. 1997).

Furthermore, harassment is
expected to be effective only in the dense regions of clusters
(i.e. virialized cores) since it depends on multiple
high-speed encounters, whereas the morphology-density relation
is shown to hold all the way from the dense core, to cluster outskirts,
and to the group environment (Dressler 1980; Dressler et al. 1997;
Cooper et al. 2007).

Moreover, harassment mechanism is expected to operate by
tearing apart the outer disks of a galaxy, or at least destorying
the spiral structure on the disk, and yet most of
the cluster post-burst population have been found to have
well-preserved disks which contain spiral structures
(Poggianti et al. 1999; Couch et al. 1998;
Goto et al. 2003b; Moran et al. 2007b).

We also want to emphasize that despite of some superficial similarities
of the harassment mechanism and tidally-enhanced
secular evolution,  the harassment simulations in fact produced very different
remnants as a result of interaction compared to interaction-enhanced secular evolution.
For example, Moore et al. (1999) 
states that for LSB galaxies ``the bound stellar remnants
closely resemble the dwarf spheroidal (dEs) that populate nearby
clusters", whereas for high surface brightness (HSB) galaxies ``although very few stars are stripped,
the scaleheight of the discs increases substantially and no spiral
features remain''.   It is clear that those dEs are not related to
cluster S0s which are the end product of BO type evolution. 
But even for the HSB galaxies, we know from the work of
Poggianti et al. (1999) and Geach et al. (2007b)
that the spiral structure and thin disks survive
the star-formation events induced by interactions.  So these remnants
cannot be the same as those from the harassment simulations, which
had their spiral structure completely destoryed by the strong tidal
shocks.  Furthermore, since secular evolution depends precisely
on the presence of density wave features (i.e., spiral arms, bars,
etc.) in the disks of galaxies to induce radial mass accretion,
the operation of tidally-enhanced secular evolution requires a
completely different galaxy-disk physical condition (cold disk with
prominent density wave structure) than what the harassment mechanism can
supply (the hot and puffed up disk devoid of density waves), despite
the superficial similarities of these two proposed mechanisms.

\subsection{Secular Evolution and the Origin of Galaxy Scaling Relations}

In the secular evolution scenario, as a galaxy evolves along the Hubble sequence
from the late to the early types, the increase in its mean
surface brightness is accompanied by a corresponding decrease
in the dynamical mass to light ratio (Broeils, Haynes, \& Baumgardt 1993;
van der Hulst et al. 1993; Zwaan et al. 1995; Bell \& de Jong 2001;
Zavala et al. 2003; and the references therein), these two effects
compensating each other to maintain the tight scaling relations
such as the Tully-Fisher and fundamental plane relations (Zhang 2004).

The empirical color-magnitude relation (CM, see, e.g. Bower,
Lucey \& Ellis 1992 and the references therein) has long puzzled astronomers
as to its origin, especially in the face of obvious morphological
transformation of galaxies as reflected in the BO effect and cluster
early-type galaxy number density evolution (Franx 2004).  The
existence of this relation in the face of morphological transformation,
however, could also naturally follow from the secular evolution process.
The aging and reddening of the stellar population
together with the gradual turning on of the star-forming 
mass reservoir (which increases the total luminosity of a galaxy)
keep the galaxies on the CM relation as they redden and get more
luminous.  The conventional wisdom had always been that 
the constant slope of CM relations across redshift z means
that this is not an age-mass relation, but rather
a metalicity-luminosity relation (Ellis et al. 1997;
Kodama et al. 1998), and the fact that most intermediate and nearby cluster
early-type galaxies fall on tight color-magnitude relation means that their stars
formed at high redshift (Franx 2004).  This, however, implicitly assumed that
the total luminosity $L$ of a galaxy does not increase with time.
When the density-wave induced secular evolution process is
considered, which results in a continued ignition of more 
and more baryonic star-forming material, as well as a continuous
radial inflow of baryons to increase the effective mass of the early
type galaxies, the total luminosity $L$ of a galaxy can increase with time. Although
with the aging and reddening of the stellar population, 
the $M/L$ ratio of the early type galaxy increases (i.e. the
average stellar population dims with age), the total
$L$ can nonetheless increase because the increase of M due to 
secular evolution is much more substantial.

Observationally, it was found that many elliptical galaxies
contain substantial outer disks (Franx 2004 and the references
therein).  The L$^*$ disky ellipticals are found to form
a continuous sequence with the S0s.  Thus the continuous
secular evolution through disk-related mass accretion
process can conceivably lead to the increase of the
elliptical galaxies' mass throughout the Hubble sequence.
For the cluster early-type population, it was found
by Fasano et al. (2000) that for clusters which have
a large elliptical population, their S0 population is correspondingly
reduced, and vice versa.  This gives indirect support to the idea
that the secular evolution process can in fact carry an
early type galaxy across the S0 Hubble type into disky ellipticals.

So far, high-z surveys have not uncovered many red cluster-galaxy
progenitors, presumably this means they are yet to be transformed
into such.  The thought-to-be high-z progenitors of the nearby
early type galaxies, the Lyman break galaxies (Steidel et al. 1996),
were found to be much less massive than the average early
type galaxies in the nearby universe (Papovich, Dickinson,
\& Steidel 2001; Shapley et al. 2001).  It is likely that
the mass of these high-z early type galaxies have grown significantly
since these earlier epochs through the secular mass accretion process.
This explanation is one way to account for the ``progenitor
bias'' as well as the increase in number density of early
type galaxies with decreasing redshift (Treu 2004 and the
references therein).

The explanation of the nature of the color-magnitude relation
as effectively an age-mass relation resolves the dilemma faced by 
the conventional explanation, that while on the one hand
the observed morphological and spectroscopic BO effect for cluster
galaxies means that not all cluster galaxy stars are old, on the other hand
the CM relation in the nearby universe is so tight which would have implied
a uniform high formation redshift of the major population of stars
in these early type cluster galaxies by the conventional wisdom.
This explanation also naturally accounts for the fact that for
every cluster observed (even those in the nearby universe), there are
always some blue outliers which were off the main track of the
cluster CM relation.  This shows that galaxies only settle onto the
CM relation as they age, since when starting off they are generally
not already on the CM relation. The CM relation is thus another ``attractor''
of the secular evolution process. 

This new explanation of the meaning of color-magnitude relation
is also consistent with the recently-established down-sizing trend
(see, e.g. Cimatti, Daddi, \& Renzini 2006; Kodama et al. 2004 and the references therein) 
for structure formation and mass assembly, which requires that the 
oldest object to be the most massive ones: this, coupled with the empirical
CM relation, means that the red objects are indeed
older, in addition to being metal rich.  The most recent observations
have by now given direct confirmation that the colors of the 
early-type galaxies in the local universe
are indeed correlated with the mean age of the
stellar population (Barr et al. 2007).

\section{Conclusions}

In this paper we have reviewed the exising dynamical mechanisms to account for
the origin of the cluster galaxy morphological transformation, and pointed out 
serious contradictions to observed
properties of cluster galaxies when many of these mechanisms are
examined in detail.  We proposed further
a new approach of using panoramic space infrared observations
to quantitatively assess the roles of the different proposed mechanisms
in transforming the cluster galaxy morphology and in producing the 
Butcher-Oemler effect. The preliminary application of this approach,
coupled with a wealth of published results analyzed under the new light,
points to interaction-enhanced secular evolution process as the
most likely contributing mechanism for producing the morphological
Butcher-Oemler effect.  This conclusion is supported, most importantly,
by the following observational evidence: that the interaction-induced
morphological transformation process in clusters appears to be a disk-related
process with well-preserved density wave patterns present on the galaxy disk
through most of the evolution process; that the transformation appears to go through a univeral
sequence of starburst, poststarburst, passive spiral, S0, and disky
elliptical stages; that the quenching of star-formation appears to
happen on a shorter timescale than the corresponding morphological
transformation timescale; that the mid-infrared signature of the interacting
events shows the predominence of a large population of moderately
infrared-bright galaxies, which have optical and near-infrared colors as well as number densities
consistent with them being the population in the transitional stage
between starburst and passive galaxies.
All of the above evidence are consistent with the interaction-enhanced
secular evolution scenario for galaxy morphological transformation in clusters.

\section*{Acknowledgments}
The author is indebted to J. Fischer, W. Couch, A. Dressler, C. Dudley, E. Dwek, E. Egami, 
G. Helou, J. Huang, A. Oemler, D. Sanders, S. Satyapal,
J. Schombert, and J. Virtelek for helpful comments and input 
to this project during the preparation of a Spitzer proposal 
to further explore the application of the IR diagnostic procedure described 
in this work.  Thanks to S. Moran for making available
the optical-spectroscopic-redshift dataset on the member galaxies
in CL 0024+16.  An anonymous refereee provided a very thorough
review of the manuscirpt, and many of his helpful suggestions are
incorporated into the current text. J. Fischer proof-read a significant
fraction of the manuscript and has made valuable comments towards its improvement.
Thanks to E. van Kampen and collaborators for 
incorporating some of the elements of the current work in
a new Herschel proposal to test the cluster galaxy evolution scenarios.
This research was supported in part by funding from the Office of Naval
Research.

\section{References}

\noindent
Abraham, R. et al. 1996, ApJ, 471, 694

\noindent
Barr, J.M., Bedregal, A.G., Aragon-Salamanca, A., Merrifield,
M.R., \& Bamford, S.P. 2007, arXiv/astroph-0705.0623

\noindent
Bershadskii, A., \& Sreenivasan, K.R. 2002, Phys. Lett. A, 299, 149

\noindent
Bershadskii, A., \& Sreenivasan, K.R. 2003, Phys. Lett. A, 319, 21

\noindent
Boselli, A., \& Gavazzi, G. 2006, PASP, 118, 517

\noindent
Bower, R.G., Kodama, T., \& Terlevich, A. 1998, MNRAS, 299, 1208

\noindent
Bower, R.G., Lucey, J.R., \& Ellis, R.S. 1992, MNRAS, 254, 601

\noindent
Butcher, H., \& Oemler, A. Jr. 1978a, ApJ, 219, 18 

\noindent
---- 1978b, ApJ, 226, 559

\noindent
Cimatti, A., Daddi, E., \& Renzini, A. 2006, A\&A, 453, L29

\noindent
Contursi, A. 1998, in Untangling Coma Berenices: A New Vision of 
an Old Cluster, eds.  Mazure, A., Casoli, F., Durret, F., Gerbal, D., p. 203

\noindent
Cooper et al. 2006, MNRAS, 370, 198

\noindent
Cooper et al. 2007, MNRAS, 376, 1445

\noindent
Couch, W.J., Matthew, M.C., \& De Propris, R. 2004, in
Clusters of Galaxies: Probe of Cosmological Structure and Galaxy
Evolution, eds. J.S. Mulchaey, A. Dressler, \& A. Oemler
(Cambridge: CUP)

\noindent
Couch, W.J., Ellis, R.S., Sharples, R., \& Smail, I. 1994, ApJ, 430, 121

\noindent
Couch, W.J. et al. 1998, ApJ, 497, 188

\noindent
Couch, W.J., \& Sharples, R.M., MNRAS, 229, 423

\noindent
Cowie, L.L., Songaila, A., Hu, E.M., \& Cohen, J.G. 1996,
AJ 112, 839

\noindent
Daddi, E., Cimatti, A., \& Renzini, A. 2000, A\&A, 362, 45

\noindent
Dale, D.A., \& Helou, G. 2002,  ApJ, 576, 159

\noindent
Dressler, A. 1980, ApJ, 236, 351

\noindent
Dressler, A. et al.  1997, ApJ, 490, 577

\noindent
Dressler, A., Smail, I., Poggianti, B.M., Butcher, H., Couch, W.J.,
Ellis, R.S., \& Oemler, A., Jr. 1999, ApJS, 122, 51

\noindent
Duc, P.-A., et al.  2002, A\&A, 382, 60

\noindent
Elbaz, D., Daddi, E., Le Borgne, D., Dickinson, M., Alexander, D.M., Chary, R.-R.,
Starck, J.-L., Brandt, W.N., Kitzbichler, M., MacDonald, E., Nonino, M., Popesso, P.,
Stern, D., \& Vanzella, E. 2007, A\&A, 468, 33

\noindent
Ellis, R.S., Smail, I., Dressler, A., Couch, W.J., Oemler, A., Jr.,
Butcher, H., \& Sharples, R.M. 1997, ApJ, 483, 582

\noindent
Evrard, A.E. 2004, in n Clusters of Galaxies: Probes of Cosmological
Structure and Galaxy Evolution, eds. J.S. Mulchaey, A. Dressler,
and A. Oemler, p.1 (Cambridge: CUP)

\noindent
Faber, S.M., \& Jackson, R.E. 1976, ApJ, 204, 668

\noindent
Fadda, D. et al. 2000, A\&A, 361, 827

\noindent
Fasano, G., Poggianti, B.M., Couch, W.J., Bettoni, D.,
Kjaerggard, P., \& Moles, M. 2000, ApJ, 542, 673

\noindent
Franx, M. 2004, in Clusters of Galaxies: Probes of Cosmological
Structure and Galaxy Evolution, eds. J.S. Mulchaey, A. Dressler,
and A. Oemler, p.196 (Cambridge: CUP)

\noindent
Fujita, Y. 2004, PASJ, 56, 29

\noindent
Fujita, Y. \& Nagashima,, M. 1999, ApJ, 516, 619

\noindent
Gamow, G., 1952, Phys. Rev. D, 86, 251

\noindent
Geach, J.E. et al. 2006, ApJ, 649, 661

\noindent
Gnedin, O.Y. 1999, in Galaxy Dynamics, eds. D. Merritt, J.A. Sellwood, 
\& M. Valluri (San Francisco: ASP), 495

\noindent
Gnedin, O.Y. 2003a, ApJ, 582, 141; 2003b, ApJ, 589, 752

\noindent
Goto, T. 2005, MNRAS, 359, 1415

\noindent
Goto, T., Okamura, S., Yagi, M., Sheth, R.K., Bahcall, N.A., Zabel, S.A.,
Crouch, M.S., Sekiguchi, M., Annis, J., Bernardi, M., Chong, S.-S.,
Gomez, P.L., Hansen, S., Kim, R.S.J., Knudson, A., McKay, T.A.,
\& Miller, C.J. 2003a, PASJ, 55, 739

\noindent
Goto, T., Okamura, S., Maki, S., Bernardi, M., Brinkmann, J., 
Gomez, P.L., Harvanek, M., Kleinman, S.J., Krzesinski, J., Long, D., 
Loveday, J., Miller, C.J., Neilsen, E.H., Newman, P.R., Nitta, A., 
Sheth, R.K., Snedden, S.A., Yamauchi, C. 2003b, PASJ, 55, 757

\noindent
Goto, T., Yamauchi, C., Fujita, Y., Okamura, S., Sekiguchi, M.,
Smail, I., Bernardi, M., Gomez, P.L. 2003c, MNRAS, 346, 601

\noindent
Goto, T., Yagi, M., Tanaka, M., Okamura, S. 2004, MNRAS, 348, 515

\noindent
Gunn, J.E., \& Gott, J.R. 1972, ApJ, 176,1

\noindent
Hansen, L. et al.  2000a, A\&A, 356, 83; 2000b, A\&A, 362, 133

\noindent
Icke, V. 1985, A\&A, 144, 115

\noindent
Illingwoth, G., Kelson, D., van Dokkum, P., \& Franx, M. 2000,
Ap\&SS, 269, 485

\noindent
Kauffmann, G. 1995, MNRAS, 274, 153

\noindent
Kauffmann, G. 1996, MNRAS, 281, 487

\noindent
Kodama, T. et al. 2004, MNRAS, 350, 1005

\noindent
Kodama, T., Arimoto, N., Barger, A.J., \& Aragon-Salamanca, A.
1998, A\&A, 334, 99

\noindent
Larson, R.B., Tinsley, B.M., \& Caldwell, C.N. 1980, ApJ, 237, 692.

\noindent
Lemonon, L., Pierre, M., Cesarsky, C.J., 
Elbaz, D. ,Pello, R.; Soucail, G.; Vigroux, L, 1998, A\&A, 334, L21

\noindent
Mann, R.G. et al. 1997, MNRAS, 289, 482

\noindent
Martini, P. et al. 2006, ApJ, 644, 116

\noindent
Mihos. J.c., \& Hernquist, L. 1994, ApJL, 431, 9

\noindent
Moore, B. 2004, in
Clusters of Galaxies: Probes of Cosmological
Structure and Galaxy Evolution, eds. J.S. Mulchaey, A. Dressler,
and A. Oemler, p.295 (Cambridge: CUP)

\noindent
Moore, B., Katz, N., Lake, G., Dressler, A., O
emler, A., Jr. 1996, Nature, 379, 613

\noindent
Moore, B., Lake, G., \& Katz, N. 1998, ApJ, 495, 139

\noindent
Moore, B., Lake, G., Quinn, T., \& Stadel, J. 1999, MNRAS, 304, 465

\noindent
Moran, S.M., Ellis, R.S., Treu, T., Smail, I., Dressler, A., Coil, A.L.,
\& Smith, G.P., 2005, ApJ, 634, 977

\noindent
Moran, S.M., Miller, N., Treu, T., Ellis, R.S., \& Smith, G.P.
2007a, ApJ, 659, 1138

\noindent
Moran, S.M., Ellis, R.S., Treu, T., Smith, G.P., Rich, R.M., \& Smail, I. 2007b,
ApJ, in press (astroph/0707.4173)

\noindent
Morris, S.L., Hutchings, J.B., Carlberg, R.G., Yee, H.K.C.,
Ellingson, E., Balogh, M.L., Abraham, R.G., Smecker-Hane,
T.A. 1998, ApJ, 507, 84

\noindent
Murphy, T.W.,Jr., Armus, L., Matthews, K., Soifer, B.T., Mazzarella, J.M.,
Shupe, D.L., Staruss, M.A., \& Neugebauer, G. 1996, AJ, 111, 1025

\noindent
Ostriker, J. 1980, Comments on Astrophysics, 8, 177

Ozernoy, L.M., 1974a,
in The Formation and Dynamics of
Galaxies, ed. J.R. Shakeshaft (IAU), 85

Ozernoy, L.M., 1974b,
in Confrontation of Cosmological
Theories with Observational Data, ed. M.S. Longair, 227

Ozernoy, L.M., 1978, in Confrontation of Cosmological Theories with
Observational Data, ed. M.S. Longair (IAU), 227

\noindent
Papovich, C., Dickinson, M., \& Ferguson, H.C. 2001, ApJ, 559, 620

\noindent
Peebles, P.J.E. 1974, ApJL, 189, 51

\noindent
Peebles, P.J.E. 1993, Principles of Physical Cosmology (Princeton:
Princeton Univ. Press)

\noindent
Poggianti, B.M., Smail, I., Dressler, A., Couch, W.J.,
Barger, A.J., Butcher, H., Ellis, R., \& Oemler, A., Jr. 1999,
ApJ, 518, 576

\noindent
Quilis, V., Moore, B., \& Bower, R. 2000, Science, 288, 1617

\noindent
Rakos, K. \& Schombert, J. 2005, AJ, 130, 1002

\noindent
Rowan-Robinson, M. et al. 1997, MNRAS, 289, 490

\noindent
Sanders, D.B., \& Mirabel, F.H. 1996, ARAA, 34, 749

\noindent
Shapley, A.E., Steidel, C.C., Adelberger, K.L., Dickinson, M.,
Giavalisco, M., \& Pettini, M. 2001, ApJ, 562, 95

\noindent
Smail, I., Dressler, A., Couch, W.J., Ellis, R.S., Oemler, A., Jr.,
Butcher, H., \& Sharples, R.M. 1997, ApJS, 110, 213

\noindent
Steidel, C.C., Giavalisco, M., Dickinson, M., \& Adelberger, K.L., AJ, 112, 353

\noindent
Tanaka, M., Goto, T., Okamurka, S., Shimasaku, K., \& Brinkman, J.
2004, in Outskirts of Galaxy Clusters: Intense Life in the
Suburbs, Proc. IAUC 195, ed. A. Diaferio, p.444

\noindent
Tran, K.-V. H., Franx, M., Illingworth, G., Kelson, D.D., i
\& van Dokkum, P. 2003, ApJ, 599, 865

\noindent
Trentham, N.A. 1997, PhDT, Univ. of Hawaii

\noindent
Treu, T. 2004, in Clusters of Galaxies: Probes of Cosmological
Structure and Galaxy Evolution, eds. J.S. Mulchaey, A. Dressler,
and A. Oemler, p.177

\noindent
Tully, R.B., \& Fisher, J.R. 1977, A\&A, 54, 661

\noindent
van Dokkum, P. et al. 1998, ApJL, 504, 17

\noindent
van Dokkum, P. \& Franx, M. 2001, ApJ, 553, 90

\noindent
von Weizsacker, C.F. 1951, ApJ, 114, 165

\noindent
Zhang, X. 1996, ApJ, 457, 125 

\noindent
Zhang, X. 1998, ApJ, 499, 93 

\noindent
Zhang, X. 1999, ApJ, 518, 613

\noindent
Zhang, X. 2003, JKAS, 36, 223

\noindent
Zhang, X. 2004, Ap\&SS, 319, 317

\noindent
Zhang, X., \& R.J. Buta, 2007a, in Proc. IAUS 235, Galaxy Evolution
across the Hubble Time, eds. F. Combes and J. Palous, p. 184

\noindent
Zhang, X., \& R.J. Buta, 2007b, AJ, 133, 2584

\noindent
\clearpage

\vfill
\eject

\end{document}